\documentclass[a4paper]{article}
\usepackage{amsmath}
\usepackage{graphicx}
\usepackage{amsfonts}
\usepackage{amssymb}
\newtheorem{theorem}{Theorem}
\newtheorem{corollary}[theorem]{Corollary}

\newenvironment{proof}[1][Proof]{\textbf{#1.} }{\ \rule{0.5em}{0.5em}}
\numberwithin{equation}{section}

\begin{document}

\title{\textbf{Duality transformations for generalized WDVV in Seiberg-Witten theory.}}
\date{}
\author{L.K. Hoevenaars}
\maketitle
\begin{abstract}
In Seiberg-Witten theory the solutions to these equations come in certain
classes according to the gauge group. We show that the duality transformations
transform solutions within a class to another solution within the same class,
by using a subset of the Picard-Fuchs equations on the Seiberg-Witten family
of Riemann surfaces. The electric-magnetic duality transformations can be
thought of as changes of a canonical homology basis on the surfaces which in
our derivation is clearly responsible for the covariance of the generalized
WDVV system.
\end{abstract}

\section{Introduction}

In 1994, Seiberg and Witten \cite{SEIB-WITT1:1994}\ solved the low energy
behaviour of pure N=2 Super-Yang-Mills theory by giving the solution of the
prepotential $\mathcal{F}$. The essential ingredients in their construction
are a family of Riemann surfaces $\Sigma$, a meromorphic differential
$\lambda_{SW}$\ on it and the definition of the prepotential in terms of
period integrals of $\lambda_{SW}$%
\begin{equation}
a_{I}=\int_{A_{I}}\lambda_{SW}\text{ \ \ \ \ \ \ \ }\frac{\partial\mathcal{F}%
}{\partial a_{I}}=\int_{B_{I}}\lambda_{SW}\label{def a,F}%
\end{equation}
The cycles $A_{I}$ and $B_{I}$ belong to a subset of a canonical homology
basis on the surface $\Sigma$ and the $a_{I}$ are the moduli parameters of the
family of surfaces. These formulae define the prepotential $\mathcal{F}\left(
a_{1},...,a_{r}\right)  $ implicitly, where $r$ denotes the rank of the gauge
group under consideration.

A link between the prepotential and the Witten-Dijkgraaf-Verlinde-Verlinde
equations \cite{WITT:1991},\cite{DIJK-VERL-VERL:1991} was proven to exist in
\cite{MARS-MIRO-MORO:1996} where it was found that the prepotential
$\mathcal{F}(a_{1},...,a_{r})$ for pure N=2 SYM theory for classical Lie
algebras (those of type $A,B,C,D$) satisfies the generalized WDVV equations
\begin{equation}
\mathcal{F}_{I}\left[  \sum_{M}\gamma_{M}\mathcal{F}_{M}\right]
^{-1}\mathcal{F}_{J}=\mathcal{F}_{J}\left[  \sum_{M}\gamma_{M}\mathcal{F}%
_{M}\right]  ^{-1}\mathcal{F}_{I}\text{ \ \ \ \ \ \ \ \ \ }\forall
I,J,K=1,...,r\label{wdvv}%
\end{equation}
where the $\mathcal{F}_{I}$ are matrices given by $\left(  \mathcal{F}%
_{I}\right)  _{JK}=\frac{\partial^{3}\mathcal{F}}{\partial a_{I}\partial
a_{J}\partial a_{K}}$ and the $\gamma_{M}$ may depend on the $a_{I}$. If
$\mathcal{F}$ satisfies $\left(  \ref{wdvv}\right)  $ for some set of
$\gamma_{M}$ it will automatically satisfy them for any other set
$\widetilde{\gamma}_{M}$ as long as $\sum_{M}\widetilde{\gamma}_{M}%
\mathcal{F}_{M}$ is invertible. These equations are indeed a generalization of
the original WDVV equations, since we no longer demand the matrix $\sum
_{M}\gamma_{M}\mathcal{F}_{M}$ to be flat and constant.

In an alternative approach to \cite{MARS-MIRO-MORO:1996}, Ito and Yang
\cite{ITO-YANG:1998} give a proof which is valid\footnote{Algebras $E_{7}$ and
$E_{8}$ are not considered in \cite{ITO-YANG:1998} because of computational
difficulty, but they are expected to follow the same pattern.} for Lie
algebras of type $A,D,E_{6}$ and it was shown in \cite{HOEV-MART:2001} that
this method can be adapted to give a proof also in the case of Lie algebras of
type $B,C$. The approach used by Ito and Yang\ consists of two main
ingredients: an associative algebra with structure constants $C_{IJ}^{K}$,
together with a relation between the structure constants and the third order
derivatives of the function $\mathcal{F}$%
\begin{equation}
\mathcal{F}_{I}=C_{I}\cdot\left(  \sum_{M=1}^{r}\gamma_{M}\mathcal{F}%
_{M}\right) \label{key relation}%
\end{equation}
for some $\gamma_{M}$. To derive this relation they used a subset of the
Picard-Fuchs equations on the Riemann surfaces. If the matrix $\sum_{M=1}%
^{r}\gamma_{M}\mathcal{F}_{M}$ is invertible then we can substitute $\left(
\ref{key relation}\right)  $ into the WDVV equations $\left(  \ref{wdvv}%
\right)  $ which then express associativity of the algebra through a relation
on the structure constants: $\left[  C_{I},C_{J}\right]  =0. $

\bigskip

In the physical context of Seiberg-Witten theory, electric-magnetic duality
transformations are very important and therefore it is a natural question if
these are symmetries of the generalized WDVV equations. The duality
transformations form a subgroup of $Sp(2r,\mathbb{Z})$ acting as linear
transformations on the vector $\left(  \overrightarrow{a},\overrightarrow
{\nabla}F\right)  $. These transformations are indeed symmetries, as was shown
in \cite{OHTA:2000} for a pure S-duality transformation and recently in
\cite{MARS-WIT:2001} for general symplectic transformations.

\bigskip

We will show in this Letter that the classes of solutions of the WDVV system
coming from Seiberg-Witten theory are invariant under duality transformations,
by giving the duality transformations their natural interpretation in the
Seiberg-Witten context. The reason for the invariance is that the Picard-Fuchs
equations leading to the relation $\left(  \ref{key relation}\right)  $ have
solutions in terms of period integrals of $\lambda_{SW}$. We require of our
solutions $\left(  \ref{def a,F}\right)  $ only that the cycles be canonical,
so they are invariant under symplectic transformations of the cycles. These
are precisely the duality transformations. In the derivation of $\left(
\ref{key relation}\right)  $\ we will rely heavily on the connection between
the Seiberg-Witten and Landau-Ginzburg theories which is quite straightforward
in the simply laced $(ADE)$ cases. For $B$ and $C$ Lie algebras this
connection is less clear \cite{HOEV-MART:2001} but we will show that one can
use the same method as in the other cases. For $F_{4}$ it is not rigorously
proven that the prepotential satisfies the WDVV equations, so we cannot say
anything definite about duality in this case\footnote{Since $G_{2}$ has rank
2, the prepotential depends only on 2 variables and it trivially satisfies the
WDVV equations.}.

\section{Duality transformations}

Duality transformations play an important role in $N=2$ Super-Yang-Mills
theory, where in the classical theory they exchange the Bianchi identity and
the equations of motion of the Yang-Mills field strength. In the quantum field
theory, these transformations were studied extensively in the context of
supergravity $\left(  \text{for a review, see \cite{WIT:2001}}\right)  $ and
they turn out to form a subgroup of $Sp(2r,\mathbb{Z})$ where $r$ is the rank
of the gauge group. The action on $\left(  a_{I},\mathcal{F}_{I}\right)  $ of
$\left(  \ref{def a,F}\right)  $ is given by
\begin{equation}
\left(
\begin{tabular}
[c]{l}%
$\overrightarrow{a}$\\
$\overrightarrow{\mathcal{F}}$%
\end{tabular}
\right)  \rightarrow\left[
\begin{tabular}
[c]{ll}%
$U$ & $Z$\\
$W$ & $V$%
\end{tabular}
\right]  \left(
\begin{tabular}
[c]{l}%
$\overrightarrow{a}$\\
$\overrightarrow{\mathcal{F}}$%
\end{tabular}
\right)  \text{ \ \ \ \ \ \ \ \ \ \ \ \ }\left[
\begin{tabular}
[c]{ll}%
$U$ & $Z$\\
$W$ & $V$%
\end{tabular}
\right]  \in Sp(2r,\mathbb{Z})
\end{equation}
There is some terminology used by physicists for some special transformations

\begin{itemize}
\item  An example of $S$-duality is a transformation for which
\[
\left(
\begin{tabular}
[c]{l}%
$\overrightarrow{a}$\\
$\overrightarrow{\mathcal{F}}$%
\end{tabular}
\right)  \rightarrow\left[
\begin{tabular}
[c]{ll}%
$0$ & $I$\\
$-I$ & $0$%
\end{tabular}
\right]  \left(
\begin{tabular}
[c]{l}%
$\overrightarrow{a}$\\
$\overrightarrow{\mathcal{F}}$%
\end{tabular}
\right)
\]
and in general $S$-duality transformations exchange the strong and weak
coupling regimes of the physical theory.

\item  An example of $T$-duality is a tranformation for which
\[
\left(
\begin{tabular}
[c]{l}%
$\overrightarrow{a}$\\
$\overrightarrow{\mathcal{F}}$%
\end{tabular}
\right)  \rightarrow\left[
\begin{tabular}
[c]{ll}%
$I$ & $0$\\
$W$ & $I$%
\end{tabular}
\right]  \left(
\begin{tabular}
[c]{l}%
$\overrightarrow{a}$\\
$\overrightarrow{\mathcal{F}}$%
\end{tabular}
\right)
\]
and in general $T$-duality transformations are perturbative, so they go from
weak coupling to weak coupling.
\end{itemize}

Seiberg and Witten used the fact that duality transformations can be thought
of as changes of a canonical basis of a family of auxiliary Riemann surfaces
in the following way. Let a canonical basis $\{A_{I},B_{I}\}$\ of homology on
a Riemann surface be given. The transformations that take this canonical basis
into another one are known to be (see e.g. \cite{FARK-KRA:1992})\ symplectic
transformations $Sp(2g,\mathbb{Z})$ acting like
\begin{equation}
\left(
\begin{tabular}
[c]{l}%
$\overrightarrow{A}$\\
$\overrightarrow{B}$%
\end{tabular}
\right)  \rightarrow\left[
\begin{tabular}
[c]{ll}%
$U$ & $Z$\\
$W$ & $V$%
\end{tabular}
\right]  \left(
\begin{tabular}
[c]{l}%
$\overrightarrow{A}$\\
$\overrightarrow{B}$%
\end{tabular}
\right) \label{transformations}%
\end{equation}
For Lie algebra $A_{r}$ the genus of the curves taken in Seiberg-Witten theory
equals the rank of the gauge group, so $\left(  \ref{transformations}\right)
$ generate transformations on $a_{I},\mathcal{F}_{I}$ since for example
\[
\int_{A_{1}+A_{2}}\lambda_{SW}=\int_{A_{1}}\lambda_{SW}+\int_{A_{2}}%
\lambda_{SW}=a_{1}+a_{2}%
\]
so
\[
\left(
\begin{tabular}
[c]{l}%
$\overrightarrow{a}$\\
$\overrightarrow{\mathcal{F}}$%
\end{tabular}
\right)  =\left(
\begin{tabular}
[c]{l}%
$\int_{\overrightarrow{A}}\lambda_{SW}$\\
$\int_{\overrightarrow{B}}\lambda_{SW}$%
\end{tabular}
\right)  \rightarrow\left(
\begin{tabular}
[c]{l}%
$\int_{U\overrightarrow{A}+Z\overrightarrow{B}}\lambda_{SW}$\\
$\int_{W\overrightarrow{A}+V\overrightarrow{B}}\lambda_{SW}$%
\end{tabular}
\right)  =\left[
\begin{tabular}
[c]{ll}%
$U$ & $Z$\\
$W$ & $V$%
\end{tabular}
\right]  \left(
\begin{tabular}
[c]{l}%
$\overrightarrow{a}$\\
$\overrightarrow{\mathcal{F}}$%
\end{tabular}
\right)
\]
In other words, electric-magnetic duality transformations in the
Seiberg-Witten context can be thought of as changes of canonical bases of
homology on families of Riemann surfaces.

For other gauge groups, the rank is always less than the genus of the curve
and we have to take a subset of the homology basis in such a way that this
subset has canonical intersection numbers $A_{I}\circ A_{J}=B_{I}\circ
B_{J}=0,$ $A_{I}\circ B_{J}=\delta_{IJ}$. If we call the linear subspace of
$H^{1}(\Sigma,\mathbb{Z})$ spanned by these cycles $X$, then duality
transformations are generated by changes of the canonical homology basis which
leave $X$ invariant.

\section{Picard-Fuchs equations and duality}

In our derivation of the generalized WDVV equations, we consider a system of
differential equations for the periods $\oint_{\Gamma}\lambda_{SW}$
(Picard-Fuchs like equations) and then substitute $a_{I}=\oint_{A_{I}}%
\lambda_{SW}$ and $\mathcal{F}_{I}=\oint_{B_{I}}\lambda_{SW}$ in them. This
way, the derivation of the WDVV equations is manifestly invariant under
changes of the canonical homology basis and therefore under electric-magnetic
duality transformations.

\bigskip

The Riemann surfaces for the $ADE$ cases read \cite{MART-WARN:1996}
\begin{equation}
z+\frac{1}{z}=W(x,u_{i})
\end{equation}
where $W$ can be though of as the one-dimensional version of the
Landau-Ginzburg superpotential. In deriving the Picard-Fuchs equations, use is
made of the flat coordinates of the corresponding Landau-Ginzburg theories.
For the $B,C$ Lie algebras, the methods of \cite{ITO-YANG:1998} are no longer
directly applicable: we have to \ `twist' the affine Lie algebra to construct
the surfaces\footnote{The $ADE$ algebras are invariant under this twisting}.
This leads to surfaces
\begin{equation}
z+\frac{1}{z}=\widetilde{W}(x,u_{i})
\end{equation}
where $\widetilde{W}$ need not have a direct relation to Landau-Ginzburg
theory. For the $B,C$ cases, where the relation with the corresponding
superpotential is still quite straightforward, it is shown in
\cite{HOEV-MART:2001} that the method of using Picard-Fuchs equations can be
adapted in order to give the WDVV equations.

In the next sections we will review the derivation of the WDVV equations in
order to see that duality transformations are symmetries of them.

\subsection{The $ADE$ cases}

The family of Riemann surfaces associated with Seiberg-Witten theory with Lie
algebras of $A,D,E$ type is given by \cite{MART-WARN:1996}
\begin{equation}
z+\frac{1}{z}=W(x,u_{i})\label{curves}%
\end{equation}
Here $W$ can be thought of as a one variable Landau-Ginzburg superpotential
for the corresponding Lie algebra. For instance, $W_{A_{r}}(x,u_{i}%
)=x^{r+1}-\sum_{i=1}^{r-1}u_{i}x^{r-i}$ is the superpotential for Lie
algebra $A_{r}$. However $W_{E_{6}}$ contains a square root of a polynomial, but
the structure constants of the chiral ring are the same as in the
three-variable situation\cite{EGUC-YANG:1997}. From the Landau-Ginzburg theory
it is known (see e.g. \cite{WARN:1993} and references therein) that we can
pass from the moduli $u_{i}$ to flat coordinates $t_{i}$ in terms of which the
Gauss-Manin connection is set to zero. The Landau-Ginzburg product structure
reads
\begin{equation}
\frac{\partial W}{\partial t_{i}}\frac{\partial W}{\partial t_{j}}=\sum
_{k=1}^{r}C_{ij}^{k}(t)\frac{\partial W}{\partial t_{k}}\frac{\partial
W}{\partial t_{r}}+Q_{ij}\frac{\partial W}{\partial x}%
\label{product structure}%
\end{equation}
which leads to the algebra
\begin{equation}
\frac{\partial W}{\partial t_{i}}\frac{\partial W}{\partial t_{j}}=\sum
_{k=1}^{r}C_{ij}^{k}(t)\frac{\partial W}{\partial t_{k}}%
\end{equation}
and the flat coordinates\footnote{The $t_{i}$ are flat coordinates for the
\emph{multivariable }Landau-Ginzburg superpotentials. In the cases of $A,D$
Lie algebras these superpotentials are respectively not and not much different
from the one variable case presented here. For the Lie algebras of type
$E_{6}$ it was shown explicitly in \cite{EGUC-YANG:1997} that the known flat
coordinates are indeed a solution to $\left(  \ref{condition flat}\right)  $
for the one-variable superpotential, and that the algebra is the same as in the
multivariable case.} satisfy
\begin{equation}
\frac{\partial Q_{ij}}{\partial x}=\frac{\partial^{2}W}{\partial t_{i}\partial
t_{j}}\label{condition flat}%
\end{equation}
Now that the Riemann surfaces $\left(  \ref{curves}\right)  $ are introduced,
we need a meromorphic differential on them
\begin{equation}
\lambda_{SW}=x\frac{dz}{z}%
\end{equation}
whose derivatives with respect to the moduli are
\begin{equation}
\frac{\partial\lambda_{SW}}{\partial t_{i}}=d\left[  ..\right]  -\frac
{\partial W}{\partial t_{i}}\frac{dx}{z-\frac{1}{z}}\label{diff lambda}%
\end{equation}
It is stated often in the literature that the forms
\[
-\frac{\partial W}{\partial t_{i}}\frac{dx}{z-\frac{1}{z}}%
\]
are holomorphic and linearly independent. For all classical Lie algebras, the
Riemann surfaces are hyperelliptic and this statement can be easily checked.
For all exceptional Lie algebras however, the Riemann surfaces are not
hyperelliptic and it is still not proven that the forms are holomorphic.

As a final ingredient for the Seiberg-Witten theory, we introduce a third set
of coordinates $a_{I}$ on the moduli space of the family of surfaces and
objects $\mathcal{F}_{I}$%
\begin{gather}
a_{I}   =\oint_{A_{I}}\lambda_{SW} \nonumber \\
\mathcal{F}_{I}  =\oint_{B_{I}}\lambda_{SW}%
\end{gather}
where we take a subset of a canonical basis for the homology. The holomorphic
parts of the differentials $\frac{\partial\lambda_{SW}}{\partial a_{I}}$ are
canonical with respect to the cycles and therefore $\frac{\partial
\mathcal{F}_{I}}{\partial a_{J}}$ is a submatrix of the period matrix, which
is symmetric. So $\mathcal{F}_{I}$ is a gradient and there exists a function
$\mathcal{F}(a_{I})$ with derivatives $\mathcal{F}_{I}$ and this socalled
prepotential solves the low energy behaviour of $N=2$ super-Yang-Mills theory.

The product structure $\left(  \ref{product structure}\right)  $ can also be
expressed in terms of the $a_{I}$ as follows
\begin{equation}
\frac{\partial W}{\partial a_{I}}\frac{\partial W}{\partial a_{J}}=\sum
_{K=1}^{r}C_{IJ}^{K}(a)\frac{\partial W}{\partial a_{K}}\left[  \sum_{M=1}%
^{r}\frac{\partial a_{M}}{\partial t_{r}}\frac{\partial W}{\partial a_{M}%
}\right]  +\widetilde{Q}_{IJ}\frac{\partial W}{\partial x}%
\end{equation}
where the structure constants are related through $C_{IJ}^{K}(a)=\sum
_{i,j,k}\frac{\partial t_{i}}{\partial a_{I}}\frac{\partial t_{j}}{\partial
a_{J}}C_{ij}^{k}(t)\frac{\partial a_{K}}{\partial t_{k}}$. Here we assume that
the transformation from $t_{i}$ to $a_{I}$ is invertible, which can be
justified: in the case of type $A$ Lie algebras, the number of moduli equals
the genus and therefore the Jacobian
\begin{equation}
\frac{\partial a_{I}}{\partial t_{j}}=\oint_{A_{I}}\frac{\partial\lambda_{SW}%
}{\partial t_{j}}\text{ \ \ \ \ }I,j=1,...,g
\end{equation}
is indeed invertible, since $\frac{\partial\lambda_{SW}}{\partial t_{j}}$ form
a basis of holomorphic forms (modulo exact forms). For the other Lie algebras,
we have to take a subset of a canonical basis. Suppose we supplement the forms
$\frac{\partial\lambda_{SW}}{\partial t_{j}}$ with more forms $\omega_{i}$ to
form a basis of holomorphic forms, then we know that the $g\times g$ matrix
\[
\left[
\begin{tabular}
[c]{llllll}%
$\oint_{A_{1}}\frac{\partial\lambda_{SW}}{\partial t_{1}}$ & ... &
$\oint_{A_{1}}\frac{\partial\lambda_{SW}}{\partial t_{r}}$ & $\oint_{A_{1}%
}\omega_{1}$ & ... & $\oint_{A_{1}}\omega_{g-r}$\\
. & . & . & . & . & .\\
. & . & . & . & . & .\\
$\oint_{A_{g}}\frac{\partial\lambda_{SW}}{\partial t_{1}}$ & ... &
$\oint_{A_{g}}\frac{\partial\lambda_{SW}}{\partial t_{r}}$ & $\oint_{A_{g}%
}\omega_{1}$ & ... & $\oint_{A_{g}}\omega_{g-r}$%
\end{tabular}
\right]
\]
has rank $g$. So the submatrix
\[
\left[
\begin{tabular}
[c]{lll}%
$\oint_{A_{1}}\frac{\partial\lambda_{SW}}{\partial t_{1}}$ & ... &
$\oint_{A_{1}}\frac{\partial\lambda_{SW}}{\partial t_{r}}$\\
. & . & .\\
. & . & .\\
$\oint_{A_{g}}\frac{\partial\lambda_{SW}}{\partial t_{1}}$ & ... &
$\oint_{A_{g}}\frac{\partial\lambda_{SW}}{\partial t_{r}}$%
\end{tabular}
\right]
\]
has rank $r$, and we can always find a square submatrix of rank $r$ by
choosing the proper cycles. Therefore we can always choose cycles in such a
way that the transformation from $t_{i}$ to $a_{I}$ is invertible. In
Seiberg-Witten theory, there is a specific prescription \cite{MART-WARN:1996}
of what cycles one has to take and invertibility of the corresponding
submatrix should be checked.

Following \cite{ITO-YANG:1998} we write down a subset of the
Picard-Fuchs\footnote{If a basis of cohomology on a family of Riemann surfaces
is given by $\{\omega_{i}\}$ where $i=1,...,2g$ then the Picard-Fuchs
equations express
\[
\frac{\partial}{\partial t_{i}}\oint_{\Gamma}\omega_{j}=\sum_{k}A_{ij}%
^{k}\oint_{\Gamma}\omega_{k}%
\]
where $\Gamma$ is some closed cycle and the $t_{i}$ are the moduli of the
family. By a subset of Picard-Fuchs equations, we mean that we can express
derivatives of some (not all) period integrals into a linear combination of
some others.} equations
\begin{equation}
\left(  \frac{\partial^{2}}{\partial t_{i}\partial t_{j}}-\sum_{k=1}^{r}%
C_{ij}^{k}(t)\frac{\partial^{2}}{\partial t_{k}\partial t_{r}}\right)
\oint_{\Gamma}\lambda_{SW}=0\label{Picard-Fuchs}%
\end{equation}
which hold for integrals along any closed cycle $\Gamma$.

We will now prove the following theorem

\begin{theorem}
The following formula holds: $\mathcal{F}_{IJK}=\sum_{L=1}^{r}C_{IJ}^{L}(a)\left[  \sum_{M=1}^{r}%
\frac{\partial a_{M}}{\partial t_{r}}\mathcal{F}_{MKL}\right]  $
\end{theorem}

\begin{proof}
Making a change of coordinates in 
$\left(  \ref{Picard-Fuchs}\right)  $ from $t_{i}$ to $a_{I}$ we get
\[
\left(  \frac{\partial a_{I}}{\partial t_{i}}\frac{\partial a_{J}}{\partial
t_{j}}\frac{\partial^{2}}{\partial a_{I}\partial a_{J}}-\sum_{k=1}^{r}%
C_{ij}^{k}(t)\frac{\partial a_{K}}{\partial t_{k}}\frac{\partial a_{M}%
}{\partial t_{r}}\frac{\partial^{2}}{\partial a_{K}\partial a_{M}}\right)
\oint_{\Gamma}\lambda_{SW}=0
\]
where we made use of the fact that the $a_{I}$ themselves are solutions to
$\left(  \ref{Picard-Fuchs}\right)  $. Now we can substitute $\Gamma=B_{K}$
and find
\begin{equation}
\frac{\partial^{3}\mathcal{F}}{\partial a_{I}\partial a_{J}\partial a_{K}%
}=\sum_{L=1}^{r}C_{IJ}^{L}(a)\left[  \sum_{M=1}^{r}\frac{\partial a_{M}%
}{\partial t_{r}}\mathcal{F}_{MKL}\right]  \label{key relation 2}%
\end{equation}
which is the relation $\left(  \ref{key relation}\right)  $ between the third
order derivatives of $\mathcal{F}(a_{I})$ and the structure constants
$C_{IJ}^{L}(a)$.
\end{proof}

To obtain the generalized WDVV equations for the function $\mathcal{F}(a_{I})
$ we need the matrix $\sum_{M=1}^{r}\frac{\partial a_{M}}{\partial t_{r}%
}\mathcal{F}_{M}$ to be invertible, which is generically true. Otherwise small
perturbations can make it so. Since the derivation of the theorem\ does not depend on the specific choice of
canonical homology basis we started with, and since duality transformations
are changes of canonical homology, we have the following

\begin{corollary}
Take a set of solutions of the WDVV system coming from Seiberg-Witten theory
with gauge group of $ADE$ type. Duality transformations leave the set of
solutions invariant.
\end{corollary}

In other words, after a duality transformation there exists a function
$\widetilde{\mathcal{F}}(\widetilde{a}_{I})$ which is in general
different\footnote{The transformations are often too difficult to perform
explicitly.} from $\mathcal{F}(a_{I})$ but
still satisfies the generalized WDVV system.

\subsection{The $B$ and $C$ cases}

Let us stresss that for $B,C$ Lie algebras, we no longer have
\[
z+\frac{1}{z}=W
\]
with $W$ a one variable version of the Landau-Ginzburg superpotential. This
will be reflected in the Picard-Fuchs equations. In fact, if we still work
with the flat coordinates of the Landau-Ginzburg superpotential $W_{BC}$ one
can derive the Picard-Fuchs equations \cite{ITO-YANG:1998}
\begin{equation}
\left(  \frac{\partial^{2}}{\partial t_{i}\partial t_{j}}-\sum_{k=1}^{r}%
C_{ij}^{k}\frac{\partial^{2}}{\partial t_{k}\partial t_{r}}-\sum_{k=1}^{r}%
\sum_{n=1}^{r}\frac{d_{n}t_{n}}{h^{\vee}}D_{ij}^{k}\frac{\partial^{2}%
}{\partial t_{k}\partial t_{n}}+\sum_{k=1}^{r}D_{ij}^{k}\frac{1}{h^{\vee}%
}\left(  1-d_{k}\right)  \frac{\partial}{\partial t_{k}}\right)  \oint
_{\Gamma}\lambda_{SW}=0\label{Pic-Fuc}%
\end{equation}
where the $C_{ij}^{k}$ are structure constants of the $W_{BC}$ Landau-Ginzburg
theory, the $d_{n}$ are so-called degrees of the Lie algebra (the exponents
+1), $D_{ij}^{k}$ depend on the $t_{i}$ and $h^{\vee}$ is the dual Coxeter
number. Making a change of coordinates to the $a_{I}$ just like we did for
$ADE$ algebras and using the fact that the $a_{I}$ satisfy $\left(
\ref{Pic-Fuc}\right)  $, we get
\begin{equation}
\left[  \frac{\partial a_{I}}{\partial t_{i}}\frac{\partial a_{J}}{\partial
t_{j}}-\sum_{k}C_{ij}^{k}\frac{\partial a_{I}}{\partial t_{k}}\frac{\partial
a_{J}}{\partial t_{r}}-\sum_{k,n}D_{ij}^{k}\frac{d_{n}t_{n}}{h^{\vee}}%
\frac{\partial a_{I}}{\partial t_{n}}\frac{\partial a_{J}}{\partial t_{k}%
}\right]  \frac{\partial^{2}}{\partial a_{I}\partial a_{J}}\oint_{\Gamma
}\lambda_{SW}=0\label{a Pic-Fuc}%
\end{equation}
These equations are not of such a form that we can derive a relation between
$\mathcal{F}_{IJK}$ and the structure constants $C_{ij}^{k}$ in the same way
as before. However, in \cite{HOEV-MART:2001} it was shown that some other
constants $\widetilde{C}_{ij}^{k}$ form structure constants of an associative
algebra and are related to $C_{ij}^{k}$ through
\begin{gather}
C_{ij}^{k}  =\widetilde{C}_{ij}^{k}-\sum_{l,n=1}^{r}D_{ij}^{l}\frac{2nt_{n}%
}{h^{\vee}}\widetilde{C}_{nl}^{k} \nonumber   \\
C_{i}  =\widetilde{C}_{i}-D_{i}\cdot\left(  \sum_{n=1}^{r}\frac{2nt_{n}%
}{h^{\vee}}\widetilde{C}_{n}\right)
\end{gather}
where the second line is in matrix form. Substituting this into $\left(
\ref{a Pic-Fuc}\right)  $ we get
\begin{gather}
 \left[  \frac{\partial a_{I}}{\partial t_{i}}\frac{\partial a_{J}}{\partial
t_{j}}-\sum_{k}\widetilde{C}_{ij}^{k}\frac{\partial a_{I}}{\partial t_{k}%
}\frac{\partial a_{J}}{\partial t_{r}}\right]  \frac{\partial^{2}}{\partial
a_{I}\partial a_{J}}\oint_{\Gamma}\lambda_{SW}+ \nonumber \\
\sum_{n}D_{ij}^{l}\frac{2nt^{n}}{h^{\vee}}\left[  \frac{\partial a_{I}%
}{\partial t_{n}}\frac{\partial a_{J}}{\partial t_{l}}-\sum_{l}\widetilde
{C}_{nl}^{k}\frac{\partial a_{I}}{\partial t_{k}}\frac{\partial a_{J}%
}{\partial t_{r}}\right]  \frac{\partial^{2}}{\partial a_{I}\partial a_{J}%
}\oint_{\Gamma}\lambda_{SW}   =0
\end{gather}
and in \cite{HOEV-MART:2001} further information about the $D_{ij}^{l}$ was
used to conclude
\begin{equation}
\left[  \frac{\partial a_{I}}{\partial t_{i}}\frac{\partial a_{J}}{\partial
t_{j}}-\sum_{k}\widetilde{C}_{ij}^{k}\frac{\partial a_{I}}{\partial t_{k}%
}\frac{\partial a_{J}}{\partial t_{r}}\right]  \frac{\partial^{2}}{\partial
a_{I}\partial a_{J}}\oint_{\Gamma}\lambda_{SW}=0
\end{equation}
From this point we can proceed with the same reasoning as in the $ADE$ case
and conclude that the following theorem holds

\begin{theorem}
The following formula holds: $\mathcal{F}_{IJK}=\sum_{L=1}^{r}\widetilde{C}_{IJ}^{L}(a)\left[  \sum
_{M=1}^{r}\frac{\partial a_{M}}{\partial t_{r}}\mathcal{F}_{MKL}\right]  $
\end{theorem}

\begin{corollary}
Take a set of solutions of the WDVV system coming from Seiberg-Witten theory
with gauge group of $B,C$ type. Duality transformations leave the set of
solutions invariant.
\end{corollary}

\section{Conclusion and outlook}

It was shown in \cite{MARS-WIT:2001} that elements of the \textit{continuous}
symplectic group $Sp(2n,\mathbb{C})$ are symmetries of the generalized WDVV
system. Because of the relation between Seiberg-Witten theory and Riemann
surfaces, this leads naturally to the study of the \textit{discrete }subgroup
$Sp(2n,\mathbb{Z})$ in the Seiberg-Witten context, and we have shown how the
interpretation of this discrete subgroup as changes of a canonical homology
basis leads automatically to the invariance of classes of solutions to the
WDVV system.

Other solutions of the generalized WDVV\ system found so far either come from
the original WDVV equations or from the context of tau functions of conformal
mappings \cite{BOYA-MARS-RUCH-WIEG-ZABR:2001}. Interpretations of the
symplectic transformations can be looked for in both contexts. This might show
whether the discrete subgroup is a natural object there as well.

Other interesting symmetries of the WDVV system have been found
\cite{DUBR:1996}\cite{MIRO-MORO:1998}, but no systematic investigation of
symmetries has been undertaken. We believe such an investigation would be interesting.

\bibliographystyle{h-physrev}
\bibliography{biblio}
\end{document}